\documentstyle[12pt]{article}
\begin{document}
\title{\bf SEMI-QUANTUM CHAOS}
\author{ Fred Cooper\thanks{Permanent address: Theoretical Division,
Los Alamos National Laboratory, Los Alamos, NM 87545. E-mail:cooper@
pion.lanl.gov},
John F.~Dawson, Dawn Meredith, and\\
Harvey Shepard\\
Department of Physics, University of New Hampshire,\\
Durham, NH 03824}
\maketitle
\begin{abstract}
We consider a system in which a classical oscillator is interacting
with a purely quantum mechanical oscillator, described by the Lagrangian
$
  L = \frac{1}{2} \dot{x}^2 + \frac{1}{2} \dot{A}^2
  - \frac{1}{2} ( m^2 + e^2 A^2) x^2 \>,
$
where $A$ is a classical variable and $x$ is a quantum operator.  With
 $\langle x(t) \rangle = 0$, the relevant variable for the
quantum oscillator is $\langle x(t) x(t) \rangle = G(t)$.  The
classical Hamiltonian dynamics governing the variables $A(t)$, $\Pi_A(t)$,
$G(t)$ and $\Pi_G(t)$ is chaotic so that the results of making
measurements on the quantum system at later times are sensitive to
initial conditions. This system arises as the zero momentum part of the problem
of pair production of charged scalar particles by a strong external electric
field.

\end{abstract}
 \section{Introduction}

The definition and observation of chaotic behavior in classical
systems is familiar and well understood \cite{bib:classicalchaos}.
However the proper definition of chaos for quantum systems and its
experimental manifestations are still unclear \cite{bib:quantumchaos}.
Here we present a simple model of a coupled quantum-classical system
and introduce a new phenomenon that we call semi-quantum chaos. In a classical
chaotic
system such as the weather we are accustomed to situations where there is lack
of long time forcasting because of the sensitivity of the system to initial
conditions. The simple model we present here has the unusual feature that
one has to give up long term forcasting even  for the quantum mechanical
probabilities,as exemplified by the average number of quanta at later times.
The complete dynamics of the coupled quantum and classical oscillators is
described  by    a
classical effective Hamiltonian that is the expectation value of the quantum
Hamiltonian. This effective Hamiltonian displays chaotic behavior, and
thus the parameters that describe the quantum mechanical wave function
(and hence expectation values) are sensitive to initial conditions.
Chaos in dynamical systems with both quantum and classical degrees of freedom
has been noted in more complicated systems and in a different context by other
authors, (see {\em e.g.}\ \cite{bib:Bonilla}).

We consider a system in which a classical oscillator is interacting
with a purely quantum mechanical oscillator described by the Lagrangian,
\begin{equation}
  L = \frac{1}{2} \dot{x}^2 + \frac{1}{2} \dot{A}^2
  - \frac{1}{2} ( m^2 + e^2 A^2) x^2 \>,
\label{eq:Lag}
\end{equation}
with equations of motion given by
\begin{eqnarray}
 \ddot{x} + ( m^2 + e^2 A^2) x & = & 0  \label{eq:xdot} \\
 \ddot{A} + e^2 x^2 \, A & = & 0 \>. \label{eq:Adot}
\end{eqnarray}
The Hamiltonian is
\begin{equation}
H = \frac{1}{2} p^2 + \frac{1}{2} \Pi_A^2 + \frac{1}{2} ( m^2 + e^2 A^2) x^2
\>,
\label{eq:Hamilt}
\end{equation}
where $p(t) = \dot{x}(t)$ and $\Pi_A = \dot{A}(t)$.
We take $x(t)$ to be a quantum operator and
$A(t)$ to be the amplitude of the classical oscillator.
We require $ [ x(t), p(t) ] = i $.
We now introduce time-independent Heisenberg representation
creation and destruction operators, $a$ and $a^{\dag}$, by the Ansatz
\begin{equation}
  x(t) = f(t) \, a + f^*(t) \, a^{\dag} \>,
\label{eq:xfa}
\end{equation}
and we note that if $f(t)$ satisfies the Wronskian condition
\begin{equation}
   i [ f^*(t) \dot{f}(t) - \dot{f}^*(t) f(t) ] = 1  \>,
\label{eq:Wron}
\end{equation}
then $a$ and $a^{\dag}$ satisfies the relation $ [ a, a^{\dag} ] = 1
$. From (\ref{eq:xdot}) and (\ref{eq:xfa}), we find that $f(t)$ satisfies
the equation of motion
\begin{equation}
   \ddot{f} + ( m^2 + e^2 A^2) f  =  0 \>,
\label{eq:fdot}
\end{equation}
with the normalization fixed by the Wronskian condition (\ref{eq:Wron}).
We can satisfy these two equations by the substitution
\[
   f(t) = \exp \left [ - i \int_0^t \Omega(t') {\rm d}t' \right ] /
       \sqrt{2 \Omega(t)} \>,
\]
where $\Omega(t)$ satisfies the nonlinear differential equation
\begin{equation}
   \frac{1}{2} \left( \frac{\ddot{\Omega}}{\Omega} \right) -
   \frac{3}{4} \left( \frac{\dot{\Omega}}{\Omega} \right)^2 + \Omega^2
    =  \omega^2 \>,
\label{eq:Omega}
\end{equation}
with
\begin{equation}
   \omega^2(t)  \equiv m^2 + e^2 A^2(t) \>.
\label{eq:omega}
\end{equation}

Now, we choose the initial state vector at $t=0$ to be the ground state of the
operator $\hat{n} = a^{\dag}a$, $| \Psi(0) \rangle = | 0 \rangle $, where
$ a | 0 \rangle = 0. $
Then, from (\ref{eq:xfa}), the average (classical) value of $x(t)$ and
$p(t)$
 is 0 for all time, $\langle x(t) \rangle = 0$ and $\langle p(t)
 \rangle = 0$.
However, the quantum fluctuations of $x(t)$ are non-zero and are given
by the variable $G(t)$,
\begin{equation}
   G(t) = \langle x^2(t) \rangle = | f(t) |^2 = \frac{1}{2 \Omega(t)}   \>.
\end{equation}
Then, from ({\ref{eq:Omega}), it is easy to show that $G(t)$ satisfies
\begin{equation}
   \frac{1}{2} \left( \frac{\ddot{G}}{G} \right) -
   \frac{1}{4} \left( \frac{\dot{G}}{G} \right)^2 -
   \frac{1}{4 G^2} + \omega^2 = 0  \>.\label{eq:Gdd}
\end{equation}
In addition, we find that
\begin{equation}
  \langle \dot{x}^2(t) \rangle = \frac{\dot{G}^2}{4 G} + \frac{1}{4 G} .
\end{equation}
The expectation value of Eq.~(\ref{eq:Hamilt}) becomes a new effective
Hamiltonian
\begin{eqnarray}
   H_{\mbox{\rm eff}} & = & \langle H(t) \rangle \nonumber \\
     & = & \frac{\Pi_A^2}{2} + 2 \Pi_G^2 G +
        \frac{1}{8 G} + \frac{1}{2} ( m^2 + e^2 A^2 ) G  \>.
\label{eq:Heff}
\end{eqnarray}
 The conjugate momenta are
\begin{equation}
\Pi_G = \frac{\dot{G}}{4 G} \>, \qquad \Pi_A = \dot{A} \>,
\label{eq:mom}
\end{equation}

This {\em classical} Hamiltonian determines the variables, $G$ and $\dot{G}$,
necessary for a complete {\em quantum-mechanical} description of this system.
Hamilton's equations then yield
\begin{eqnarray}
\dot{\Pi}_G  & = & - 2 \Pi_G^2 + {1\over 8 G^2 } -
         \frac{1}{2} \omega^2 \nonumber \\
\dot{\Pi}_A  & = & - e^2 A G
\end{eqnarray}
or equivalently:
\begin{eqnarray}
\ddot{A} + e^2 G \, A & = & 0 \nonumber \\
\frac{1}{2} \left( \frac{\ddot{G}}{G} \right) -
   \frac{1}{4} \left( \frac{\dot{G}}{G} \right)^2 -
   \frac{1}{4 G^2} + \omega^2  & = & 0  \>,
\label{eq:classeq}
\end{eqnarray}
which correspond to (\ref{eq:Gdd}) and the
expectation values of Eq.~(\ref{eq:Adot}).

The classical effective Lagrangian is
\begin{equation}
   L_{\mbox{\rm eff}}  = \frac{\dot{A}^2}{2} + \frac{\dot{G}^2}{8 G} -
        \frac{1}{8 G} - \frac{1}{2} ( m^2 + e^2 A^2 ) G  \>.
\end{equation}
This Lagrangian could also have been obtained using Dirac's action,
\begin{equation}
\Gamma =\int dt \langle\Psi(t) | i {\partial \over \partial t} - H |\Psi(t)
\rangle \equiv \int dt \, L_{\rm eff},
\label{eq:Leff}
\end{equation}
and a time-dependent Gaussian trial wave function as described
in \cite{bib:CPS}. This variational method was used to study the quantum
Henon-Heiles problem  in a mean-field approximation
\cite{bib:Patt}.
The Gaussian trial wave function is parametrized as follows
\[
\Psi(t)= [2 \pi G(t)]^{-1/4} \exp[ -(x-q(t))^2
( G^{-1}(t)/4 - i \Pi_G(t))+ ip(t)(x-q(t))] .
\]
We see that $G(t)$ and $\Pi_G(t)$ are the time dependent real and imaginary
parts
of the width of the wave function.
One can prove for our problem that if the  quantum oscillator starts at $t=0$
as
a Gaussian, it is described at all times by the above expression, where $G(t)$
and $\Pi_G(t)$ are totally determined by solving the effective Hamiltonian
dynamics. (For our special initial conditions $p(t)=q(t)=0$). Thus we find that
our effective Hamiltonian totally determines the time evolution of the quantum
oscillator.
    One interesting ``classical'' variable is the expectation value of the time
dependent adiabatic number operator, which corresponds to the number of quanta
in a situation where the classical $A$ field is changing slowly
(adiabatically).  For the  related field theory problem (see Section
3) of pair production of charged pairs by strong electric fields, this
corresponds to the time dependent single particle distribution function of
secondaries.
To find the expression for the  number of quanta, we begin with the
wave function corresponding to a slowly varying classical
field $A$:

 \[
   g(t) = \exp \Big[ - i \int_0^t \omega(t') {\rm d}t' \Big] \Big/
       \sqrt{2 \omega(t)} ,
\]
in terms of which we can decompose the quantum operator via
\begin{equation}
    x(t)  =  g(t) \, b(t) +
        g^{\ast}(t) \, b^{\dag}(t) \>.
\end{equation}
Requiring  the momentum operator to have
the form \[
    p(t)  = \dot{x}(t)
          =  \dot{g}(t) \, b(t) +
        \dot{g}^{\ast}(t) \, b^{\dag}(t)
\]
by imposing
$ g(t) \dot{b}(t) + g^{\ast}(t) \dot{b}^{\dag}(t) = 0 $,
and recognizing that $g(t)$ and $g^{\ast}(t)$  satisfy the
Wronskian condition by construction, then $b(t)$ and $b^{\dag}(t)$ have
the usual interpretation as creation and annihilation operators. That is,
$[x(t),p(t)] = i$ and $ [ b(t), b^{\dag}(t) ] = 1 $. Also
 \[
   b(t)  =  i [ g^{\ast}(t) \,
       \dot{x}(t) -
      \dot{ g}^{\ast}(t) \,  x(t) ]  \>.
\]
$b^{\dag}(t) b(t) $ can be
interpreted as a {\em time-dependent} number operator for a slowly varying
(adiabatic) classical field $A$\/. The time independent basis and the time
dependent basis are both complete sets and are related by a unitary
Bogoliubov transformation,
$ b(t) = \alpha(t) \, a + \beta(t) \, a^{\dag} $,
where
\begin{eqnarray*}
   \alpha(t) & = &
      i [ g^{\ast}(t) \dot{f}(t) -  \dot{g}^{\ast}(t) f(t) ]
      \\
   \beta(t) & = &
      i [ g^{\ast}(t) \dot{f}^{\ast}(t) - \dot{g}^{\ast}(t) f^{\ast}(t) ]
      \>,
\end{eqnarray*}
and where $ |\alpha(t)|^2 - |\beta(t)|^2 = 1 $.
 If we choose for initial conditions,
$\Omega(0) = \omega(0)$, $\dot{\Omega}(0) = \dot{\omega}(0)$,
then one finds that $\alpha(0) = 1$ and $\beta(0) = 0$.
These are the initial conditions appropriate to the field theory problem of
pair production.
The average value of the time-dependent occupation number is given by

\begin{equation}
   n(t) = \langle \, b^{\dag}(t) b(t) \, \rangle  =
   |\beta(t)|^2 =           (4 \Omega \omega )^{-1}
      \left[ (\Omega - \omega)^2 +
           \frac{1}{4} \left({\dot{\Omega} \over \Omega} -
               { \dot{\omega} \over \omega} \right)^2  \right]  \,
    \>.
\label{eq:aven}
\end{equation}

Eq.~(\ref{eq:aven}) allows us to compute the average occupation
number of the system as a function of time.

\section{Solutions of the classical equations}

We first scale out the mass by letting $ t \rightarrow m^{-1} \, t $,
$ A \rightarrow m^{-1/2} \, A $, $G \rightarrow m^{-1} \, G $, and
$ e \rightarrow e \, m^{3/2}$.  Then
 the scaled equations of motion are
\begin{eqnarray}
\ddot{A} + e^2 G \, A & = & 0 \nonumber \\
\frac{1}{2} \left( \frac{\ddot{G}}{G} \right) -
   \frac{1}{4} \left( \frac{\dot{G}}{G} \right)^2 -
   \frac{1}{4 G^2} + 1 + e^2 A^2  & = & 0  \>.
\label{eq:classeqscaled}
\end{eqnarray}

In order to explore the degree of chaos as a function of (scaled) energy
and coupling parameter $e$, we calculated surfaces of section and
Lyapunov exponents.  The surface of section is a slice through the
three-dimensional energy shell \cite{bib:classicalchaos}. That is, for a
fixed energy and coupling parameter the points on the surface of
section are generated as the trajectory pierces a fixed place (e.g.
$A=0$) in a fixed direction.  The hallmark of regular motion is the
cross section of a KAM torus which is seen as a closed curve in the surface
of section.  The hallmark of chaotic motion is the lack of any such
pattern in the surface of section.  In Fig. 1 we show a
plot of a surface of section at $E=0.8$ and $e=1.$ where regular and
chaotic regions co-exist.

The Lyapunov exponent provides a more quantitative, objective measure
of the degree of chaos.  The Lyapunov exponent, $\lambda$,
gives the rate of exponential divergence of infinitesimally close
trajectories \cite{bib:lyapunov}. Although there are as many
 Lyapunov exponents as degrees of
freedom, it is common to simply give  the largest of
these. For regular trajectories $\lambda =0$; for chaotic trajectories
the exponent is positive.
We define
\begin{equation}
\vec\eta(t) \equiv \lim_{\vert \vec \delta \vert \rightarrow 0}
{ \vec z (\vec z_0 + \vec \delta, t) -
\vec z (\vec z_0 , t) \over \vert \vec \delta \vert} \>,
\label{eq:eta}
\end{equation}
where $\vec z (\vec z_0 , t) $ is a point in phase space at time $t$
with initial position $\vec z_0$.  Then the time evolution for $\vec
\eta(t)$ is
\begin{equation}
\vec {\dot \eta}(t)  = \vec \eta(t) \cdot \vec \bigtriangledown
\vec F \mid_{\vec
z(\vec z_0, t)} \>,
\label{eq:etatime}
\end{equation}
where
\begin{equation}
\vec{\dot z} (t) = \vec F (\vec z(t),t)
\end{equation}
are the full equations of motion for the system.
The Lyapunov exponent is defined as
\begin{equation}
\lambda \equiv \lim_{t \rightarrow \infty} { 1 \over t} \ln \left \vert {
\vec \eta(t) \over \vec \eta (0) } \right \vert \>.
\label{eq:lyapunov}
\end{equation}
Appendix {\em A} of Ref.~\cite{bib:lyapunov} provides an explicit algorithm for
the calculation of all the Lyapunov exponents. Since we cannot carry out the $t
\rightarrow \infty$ limit computationally, the regular trajectories are those
for which $\lambda(t)$ decreases as $1/t$, while the chaotic trajectories give
rise to $\lambda(t)$ that is roughly constant in time, as judged by a linear
least-squares fit of $\log[\lambda(t)]$ {\em vs.}\ $\log(t)$.

We calculated the Lyapunov exponents for three values of the scaled coupling
constant $e$ ($0.1$, $1.0$, $10.0$) and for energies from $0.5$ to $2.0$.
$E=0.5$ is
the lowest energy possible, corresponding to the zero point energy of the
oscillator; there is no upper limit on $E$.  Fifty initial conditions
were chosen at random for each energy bin of width $0.1$ and coupling
parameter.  One relevant quantity to study is the chaotic volume, the fraction
of initial conditions with positive definite Lyapunov exponents
(corresponding to chaotic behavior).
Errors in this quantity arise because of the finite number of initial
conditions chosen,
and because the distinction between zero and positive exponents cannot
be made with certainty at finite times.  We found that for $e=0.1$,
more than $95\%$ of trajectories were regular for all energies tested;
for $e=1.0$ and $10.0$, there is a steadily increasing fraction of chaotic
orbits between $0.5 \le E \le 1.25$.  For $1.25 \le E \le 2.0$, more than
$90\%$ of these orbits are chaotic.

\section{Interpretation}

We may now ask what are the physical ramifications of our results.  The
system of equations studied here is the $k=0$ mode contribution to the
problem of pair production of charged mesons by a strong electric field
\cite{bib:coopetal}, with $E(t) = - \dot{A}(t)$ being the value of the
time evolving electric field.   For that problem the equation for $G(t)$
gets modified and becomes a function of the momentum $k$ of the normal
modes of the charged scalar field.
Eq.~(\ref{eq:omega}) being replaced by
$\omega^2(t) \rightarrow \omega_k^2 (t) = (k-eA(t))^2 + m^2$.

The semi-classical equation for $A(t)$ becomes
\begin{equation}
\ddot{A}=  e \int dk (k-eA(t)) G_k(t) \>,
\end{equation}
which gets contributions from all modes.  This system of equations is
discussed in detail in \cite{bib:coopetal}.  When we sum
over all the $k$ modes, $A(t)$ becomes a smooth function of time
and is insensitive to initial data.  However the number of particles
produced in a narrow bin of momentum between $k$ and $k+dk$ depends
only on $G_k(t)$ and $\dot{G}_k(t)$.  Only if one does a coarse
graining over momentum does one lose this sensitivity to initial data.
Thus one should observe, as one counts the number of produced charged
particles in a detector and increases the resolution, that the number
of counts in a narrow momentum bin becomes a rapidly oscillating
function of time whose behavior is chaotic. The expression for the
number of particles in a given momentum bin $k$ is
\begin{equation}
n(k,t) = \left(4 \Omega_k \omega_k \right)^{-1} \left( (\Omega_k
-\omega_k)^2 + {1 \over 4} \left({\dot{\Omega}_k  \over \Omega_k}
-{\dot{\omega}_k  \over \omega_k} \right)^2 \right),
\end{equation}
which should be compared with Eq.~(\ref{eq:aven}).
The chaotic behavior of Eq.~(\ref{eq:aven}) is shown in Fig. 2.
[Also see  Fig.~3 of
\cite{bib:coopetal}].

\vspace{1.0in}

\noindent{\bf FIGURE CAPTIONS}\\
\vspace{0.2in}

\noindent Figure 1: A plot of the surface of section for energy = $0.8$,
 $e=1.0$, and $A=0$.
Each symbol represents a different trajectory. The one chaotic
region is in the center of the plot.

\noindent Figure 2:  A plot of the occupation number given by
Eq.~(\ref{eq:aven}) for energy~=~ $1.8$, $e=1.0$, $A(0)=0$, $\Pi_G(0)=0$.
The solid line is for for $G(0) = 0.5$; the dashed line is for $G(0) =0.5001$.
This plot shows the sensitivity to initial conditions.

\end{document}